\documentstyle[aps,prl,epsf,twocolumn]{revtex}
\draft
\input epsf
\epsfverbosetrue
\def\be{\begin{equation}}
\def\ee{\end{equation}}

\begin{document}
\title{Conformations of Linear DNA}
\author{Boris Fain and Joseph Rudnick}
\address{Department of Physics, University of California, Los
Angeles CA 90095, USA}
\author {Stellan \"Ostlund}
\address{Institue for Theoretical Physics, Chalmers Institute of Technology,
Gothenberg, Sweden}  
\date{\today}
\maketitle
\begin{abstract}
We examine the conformations of a model for under- and overwound
DNA. The molecule
is represented as a cylindrically symmetric elastic string subjected
to a stretching force and to constraints corresponding to a
specification of the link number. We derive a fundamental relation between
the Euler angles that describe the curve and the topological linking number. Analytical expressions
for the spatial configurations of the molecule in the infinite
length limit are obtained. A unique
configuration minimizes the energy for a given set of physical conditions. An elastic model incorporating thermal fluctuations provides excellent agreement 
with experimental results on the plectonemic transition.
\end{abstract}

\section{History and Introduction}

Conformations of a slender elastic rod were originally viewed as an
interesting
problem in classical elasticity theory. Kirchhoff \cite{kirchhoff}
 was the first to make
significant headway towards a complete solution. Almost a century later,
as polymers became the subject of intense study, interest in the
problem picked up once again\cite{vologod}.
After the discovery of biological polymers---nucleic
acids and proteins---researchers recognized the importance of predicting the
elastic shape of linear molecules. The shape (``tertiary structure'')
of DNA and RNA plays an important role in the processes of replication
and transcription. Because of this a number of authors have  analyzed
various aspects of elastic DNA
conformation \cite{crick,fuller} for both closed (circular)
\cite{lebret,tsu,tanak} and
open (linear) \cite{benham} configurations. The approaches taken include
Lagrangian
mechanics
\cite{lebret,benham,tsu,tanak}, (numerical) molecular dynamics \cite{olsen}
and statistical mechanics\cite{siggia,hunt}. Despite significant progress
\cite{benham}, the equilibrium configurations of infinitely long open DNA
have not
been
analytically described. Our main aims are to set up a formalism for obtaining 
equilibrium configurations; to find one such conformation for stretched twisted 
DNA, and to set up a model of plectonemic transition to compare with 
experimental results \cite{strick}.

The work reported here is based on two major results obtained over a century
apart.
Kirchhoff \cite{kirchhoff} provided the basic framework of elasticity
theory. He observed that the
equations of equilibrium that
describe an elastic rod
are formally identical to the equations of motion of a heavy symmetric top
with one point fixed (see figure \ref{duality_fig}).
The rod Hamiltonian is identical to the top Lagrangian,
with arclength mapping onto time. This duality provides an added insight
into the nature of the solutions.
White's Theorem \cite{white} provides another crucial analytical tool. This
important theorem relates
the linking
number---a natural topological invariant of strings and ribbons---to two
manageable
components, each calculable in terms of locally defined quantities.
According to White's theorem, ${\cal L}k = {\cal T}w  +  {\cal W}r$. ${\cal
L}k$ is the linking number, while
${\cal T}w$, the twist, monitors the twist of the molecule about its axis, and
${\cal W}r$, the writhe, records the contortions of the axis. This theorem
greatly simplifies the problem of formulating a constraint on the linking
number. In addition we make use of the results of Fixman and Kovac 
\cite{fixman} and Marko and Siggia\cite{siggia,siggia_comment} to develop 
thermal effects of the plectonemic transition. 

\section{The Model}

The elastic model of DNA represents the molecule as a slender cylindrical
elastic rod. To model external forces and torques
the rod is stretched (in the $z$ direction) by a force $F$ and is required to
have a fixed ${\cal L}k$ (see figure \ref{model_fig} top). The rod is
parametrized by arclengh,
$s$. At each point $s$ we describe the rod
by relating the local coordinate frame ${\cal L}$ to the frame ${\cal L}_0$
rigidly embedded in the curve in its relaxed configuration. 
The relationship between the stressed and
unstressed local frames is specified by Euler angles $\theta(s),
\phi(s), \psi(s)$ needed to rotate ${\cal L}_0$ into $\cal L$ (see
figure \ref{model_fig} bottom).
The shape of the backbone ${\bf r}(s)$ is traced out by the unit tangent
${\bf t}(s)$.
${\bf n}(s)$, a unit normal, keeps track of the twist, ${\cal T}w$.
In this paper we will often omit $s$-dependence for brevity. We also make
use of the notation
$\dot{x}\equiv\frac{d}{ds}(x)$. Then,
\begin{eqnarray}
{\bf r}(s)&=&\int_{0}^{s}{\bf t}(s')\;ds'  \label{r_eqn}\\
{\bf t}(s)& =&
\left(\sin{\theta}\sin{\phi},\,\sin{\theta}\cos{\phi},\,\cos{\theta}\right)
\label{t_eqn}\\
{\bf n}(s) &=&
(\cos\phi\cos\psi - \cos\theta\sin\phi\sin\psi,  \nonumber \\
& &\,-\cos\phi\sin\psi - \cos\theta\cos\phi\sin\psi,\,
\sin\theta\sin\psi )
\label{n_eqn}
\end{eqnarray}

\noindent Let the elastic constants of bending and torsional stiffness be
denoted,
respectively, by  $A$ and $C$, and let $L$ be the lengh of the
rod. The energy of the twisted, stressed rod is the sum of bending and
twisting energies and the potential energy produced by the stretching 
force $F$. Using (\ref{t_eqn}) and (\ref{n_eqn})): 
\begin{eqnarray}
E_{tot}&=&E_{el}-F\cos\theta=E_{bend}+E_{twist} - F\cos\theta= \nonumber  \\
&=&\frac{A}{2}\int_{0}^{L}\left({\bf \dot{t}} \right)^2\;ds + 
\frac{C}{2}\int_{0}^{L}
\left(\left({\bf n}\times\dot{\bf n}\right)\cdot{\bf t}\right)^2\;ds + \nonumber \\
&-&F\cos\theta\; \label{energy_temp} 
\end{eqnarray}
Using (\ref{t_eqn}) and (\ref{n_eqn}) in (\ref{energy_temp}) we obtain 
\begin{eqnarray}
E_{tot}&=&\int_{0}^{L}ds\;\; \,\,
\frac{A}{2}\left(\dot{\phi}^2\sin^2\theta+\dot{\theta}^2\right) +\nonumber\\
&&\frac{C}{2}\left(\dot{\phi}\cos\theta+\dot{\psi}\right)^2 
-F\cos\theta .\label{e_eqn}
\end{eqnarray}

The feature that sets this work apart from previous attempts and
allows
us to unambiguously determine a unique configuration for a given set of
initial conditions is the constraint of maintaining a fixed linking
number, ${\cal L}k$. Although linking number is usually associated 
with closed curves, the bound ends of our string allow us to define a fractional linking number for it. A caveat is that the local expressions we derive 
are only valid for the configuration (extended) we consider. White's theorem \cite{white} allows us to express
${\cal L}k$
in terms of its components, ${\cal T}w$ and ${\cal W}r$. Using (\ref{t_eqn})
and(\ref{n_eqn}):
\begin{eqnarray}
{\cal T}w&\equiv&
\frac{1}{2\pi}\int_{0}^{L}\left({\bf n}\times\dot{\bf n}\right)\cdot{\bf t}\;ds=
-\frac{1}{2\pi}\int_{0}^{L}\dot{\phi}\cos\theta+\dot{\psi}\;ds \label{tw_eqn}\\
{\cal W}r&\equiv& \frac{1}{4\pi}\int_{0}^{L}ds\int_{0}^{L}ds'\;
\frac{\left({\bf r}-{\bf r'}\right)\cdot\left({\bf t}\times{\bf t'}\right)}
{\left|{\bf r}-{\bf r'}\right|^{3}} \label{wr_eqn1}
\end{eqnarray}
A local expression for twist follows straightforwardly, as evidenced by the
far right hand side of Eq. (\ref{tw_eqn}).
The writhe, however, is not yet suitable for use as a Lagrange multiplier.
To express it as
an integral of a local quantity we use a theorem by Fuller\cite{fuller}.
The theorem allows
us to define ${\cal W}r$ locally using a diffeomorphism onto a {\em
reference} curve,
${\cal C}_r$. (Once again, we must stress that a different configuration, 
e.g. a circular plasmid, would require a different reference curve, 
producing slightly modified local expressions). 
The writhe is expressed as
\be
{\cal W}r={\cal W}r_r+\frac{1}{2\pi}\int_{0}^{L}
\frac{{\bf t}_r\times{\bf t}\cdot\frac{d}{ds}\left({\bf t}_r+{\bf t}\right)}
{1+{\bf t}_r\cdot{\bf t}}\;ds\,\,, \label{fuller_thm}
\ee
where ${\cal W}r_r$ is the writhe of the reference curve. Finding a suitable
reference curve proves crucial. The best choice is also the
simplest - a straight line ${\cal C}_r=\left(0,0,s\right)$.
This gives ${\bf t}_r=(0,0,1)$ and ${\cal W}r_r=0$. Substituting
(\ref{t_eqn}) into
(\ref{fuller_thm})
we obtain
\be
{\cal
W}r=\frac{1}{2\pi}\int_{0}^{L}\;\dot{\phi}\left(\cos\theta-1\right)
\;ds
\label{wr_eqn}
\ee
Combining (\ref{tw_eqn}) and (\ref{wr_eqn}) we are led to the simple
expression for ${\cal L}k$:
\be
{\cal L}k\;=\;-\frac{1}{2\pi}\int_{0}^{L}\left(\dot{\phi}
+\dot{\psi}\right)\;ds \label{lk_eqn}
\ee
\noindent Thus we have derived a simple conservation law that expresses
the invariance of  ${\cal L}k$.
\noindent Inserting $p\,d{\cal L}k$ into the right hand side of Eq.
(\ref{e_eqn}), with $p$ a Lagrange multiplier,
the expression to be minimized becomes
\begin{eqnarray}
{\cal H} &&= \int_{0}^{L}\;
\frac{A}{2}\left(\dot{\phi}^2\sin^2\theta+\dot{\theta}^2\right) \nonumber \\
&&+\frac{C}{2}\left(\dot{\phi}\cos\theta+\dot{\psi}\right)^2
-F\cos\theta -p\left(\dot{\phi}
+\dot{\psi}\right)\;ds \label{hamiltonian}
\end{eqnarray}
DNA conformations of minimum energy are found among the extrema of $\cal H$.

\section{Solutions}
We find the extrema of $\cal H$ by applying standard variational techniques
to (\ref{hamiltonian}). The resulting Euler - Lagrange equations for
$\theta(s)$, $\phi(s)$ and $\psi(s)$ are:

\begin{eqnarray}
\dot{\phi} &= & \frac{p\left(1-\cos\theta\right)}{A\sin^{2}\theta}
\hspace{.2in}
\dot{\psi} =
\frac{p}{C}\;-\;\frac{p\left(1-\cos\theta\right)\cos\theta}{A\sin^{
2}\theta}
\label{angles_eqn}\\
\frac{A}{2}\dot{\theta}^2 &= &-\;\frac{p^2}{A\left(1+\cos\theta\right)}\;-\;
F\cos\theta\;+\;E_0 \label{theta_eqn}
\end{eqnarray}

A central goal is to find a unique conformation of the rod given
a set of externally imposed constraints, $F$ and ${\cal L}k/L$, where $L$ is
the length of the rod.
We find that equations (\ref{theta_eqn}) and (\ref{angles_eqn})
support two types of solutions. The first is a family of twisted vertical
lines:
\be
\theta=0 ; \hspace{0.3in} \phi=0 ; \hspace{0.3in}
\psi(s)=\left(2 \pi {\cal L}k/L\right)\;s
\ee
The energy of the straight line follows directly from (\ref{e_eqn}):
\be
E_{line}=\frac{2CT}{2}\left(\pi {\cal L}k/L\right)^2
\ee

The second family of
solutions can be extracted from (\ref{theta_eqn}).
Multiplying Eq. (\ref{theta_eqn}) by $\sin^2\theta$
and integrating once we can rewrite (\ref{theta_eqn}) in the
following form ($u\equiv\cos\theta$):
\begin{eqnarray}
ds&=&\frac{du}{\sqrt{-\,\frac{2p^2}{{A}^2}\left(1-u\right)+\frac{2}{A}
\left(E_{0}-Fu\right)\left(1-u^{2}\right)}} \nonumber \\
&\equiv&
\frac{du}{\sqrt{\frac{2F}{A}
\left(u-a\right)\left(u-b\right)\left(u-c\right)}}\hspace{.3in}
\left(a\leq b\leq c\right) \label{u_eqn}
\end{eqnarray}
These ``writhing'' solutions are characterized by the roots $\{a,b,c\}$ of the
cubic
polynomial in the denominator of (\ref{u_eqn}) (see figure \ref{poly_fig}).
One of the roots, either $a$ or $b$, is $1$. If $u=1$ is a single root, then
the configurations form ``superhelices.'' If $u=1$ is a double root
($a=b=1$) the molecule supports a soliton-like excitation (See figure
\ref{energy_fig}). The quantity $u=\cos\theta$ takes on values between $c$
and $b$,
the quadrature turning points. The expressions for $\phi$, $\psi$
and other quantities of interest follow by quadratures. The integrals
are easily evaluated in terms of elliptic functions.

We have investigated the properties of solutions to (\ref{angles_eqn}) and
(\ref{theta_eqn}), with the bending and torsional stiffness, $A$ and
$C$, appropriate to DNA \cite{stiff1,tsu}.  Using both numerical and
analytical methods we find that, given a
particular ${\cal L}k/T$, the member of the writhing family with the
lowest energy is the soliton configuration ($a=b=1$) (see figures
\ref{poly_fig} and \ref{energy_fig}). The shape of this solution is
defined by the following relationship between the arclengh $s$ and
$u\equiv\cos\theta$:
\be
s\left(u\right)= \sqrt{\frac{2F}{A\left(1-c\right)}}
\ln\left(\frac{1+\sqrt{\frac{u-c}{1-c}}}{1-\sqrt{\frac{u-c}{1-c}}}  \right)
\label{soliton_s_eqn}
\ee
where the lower root, $c$, is given by
\begin{eqnarray}
c&=&\frac{2\left(\pi C {\cal L}k/L\right)^2}{AF}-1 \nonumber \\
&=& \frac{p^2}{4AF}-1 ,
\label{soliton_s_eqn1}
\end{eqnarray}
and $\phi$ and $\psi$ are similarly determined. A very interesting quantity
is the energy of the soliton and its relationship to that
of the twisted line solution (which are both infinite in the limit
$L \rightarrow \infty$):
\begin{eqnarray}
E_{soliton}=E_{line}+\Delta E \hspace{.2in} \mbox{with} \;\Delta E = \nonumber \\   
\frac{4F}{L}\sqrt{1+c}\left[\sqrt{\frac{1-c}{1+c}} - \arctan \left( \sqrt{\frac{1-c}{1+c}}\right) \right] \geq 0
\label{soliton_e_eqn}
\end{eqnarray}
It's clear from (\ref{soliton_e_eqn}) that the $E_{soliton}$, while smaller
than
the energies of the other members of the ``writhing'' family, is always
{\em greater} (by a {\em finite} amount) than
the energy of the twisting line configuration satisfying the same conditions.

\section{Plectonemic Transition}

Thus we find that an extended solution that minimizes energy and has a specified
${\cal L}k/L$ it is {\em always} a twisted straight line. To check
stability we perturb the straight line solution
$\theta(s)=0+\delta\theta(s)$. The perturbation calculation shows that
the non-trivial zero-energy mode satisfies
\be
\left(A\frac{d^2}{dt^2}-\left(F-\frac{p^2}{4A}\right)\right)
\delta\theta(s)=0 \label{perturb_eqn}
\ee
Thus for
${\cal L}k/L\leq\frac{A}{I_3}\pi\sqrt{F}$
the straight line $\theta=0$ is stable to small fluctuations. What happens to the molecule when
the ${\cal L}k/L$ approaches the critical value? Our results strongly indicate that the molecule attempts to loop 
over and pass through itself to shed a unit of ${\cal L}k/L$ and thus 
starts to form a plectonemic bubble. In this sense the twisted rod is in a metastable state. The plectoneme plays the role of the ``bounce'' via which a system tunnels out of the false vacuum. Beyond the transition to local instability the plectonemes ought to proliferate. To explore this scenario we 
formulate a very simple model of the plectonemic transition of stretched 
twisted DNA and compare its predictions with recent beautiful experiments 
by Strick et al.\cite{strick}
\subsection{plectonemic transition model} 

The model is diagrammed in fig. \ref{plect_fig}. In the following all 
quantities are normalized by the lengh $L$. DNA researchers prefer 
to use $\sigma \equiv \Delta{\cal L}k_{tot}/{\cal L}k_{0}$ to measure topological properties of DNA. Here 
${\cal L}k_{0}$ is the natural link of the unstressed DNA molecule; 
B-DNA has one right-handed twist every $h=3.4 nm$.  
 We will follow this notation.

The molecule is constrained 
to have a total Link ${\cal L}k_{tot} = \sigma_{tot}/h$.   
The plectonemic fraction takes up $X$ 
leaving $1-X$ straight. The plectoneme 
has a radius $R$ and a pitch $P$. The straight portion is twisted to its 
critical value $d{\cal T}w = h\sqrt{AF}/\left(2\pi C\right)
 \equiv \sigma_{l}/h$. The actual twist is slightly below critical\cite{soc},
 but numerical results indicate that the precise value 
(which depends weakly on $L$) is adequately approximated 
by that of an infinite string\cite{Boris}.  Guided by ``twist conservation'' implied by 
eqns. (\ref{angles_eqn}) we assign the same rate of $d{\cal T}w$ 
to the plectoneme. The remaining link,
${\cal L}k_{tot}- \left({\cal T}w_{pl}+{\cal L}k_{l}\right)$, is absorbed by the plectoneme's ${\cal W}r_{pl}$. Let us account the link distribution: 
\begin{eqnarray}
{\cal L}k_{l}&=&{\cal T}w_{k}=\frac{\sigma_{l}}{h}X \hspace{.15in}  
\mbox{and} \hspace{.15in}{\cal L}k_{p} = {\cal L}k_{tot}-{\cal L}k_{l}= \nonumber \\
&=&\frac{\sigma_{tot}-\sigma_{l}}{h}
+\frac{\sigma_{l}}{h}X = {\cal W}r_{p}+{\cal T}w_{p}
\label{plect_inv_eqn}
\end{eqnarray}
Because the plectoneme has the same rate of twist as the line, we 
can read off its writhe from eqn. (\ref{plect_inv_eqn}). At the same time
the writhe of a plectoneme can be expressed as a function of $P$ 
and $R$ \cite{siggia_science,white_book}. This gives us a constraint:  
\be
\label{plect_constr_eqn}
\frac{\sigma_{tot}-\sigma_{l}}{h}={\cal W}r_{p}=\frac{XP}
{2\pi\left(R^2+P^2\right)}
\ee

Up to now we have not considered any thermal effects or 
corrections. Our aim is to build a formalism of obtaining 
equilibrium zero-temperature solutions about which a thermodynamic 
theory can be obtained (e.g. by considering fluctuations)
\cite{siggia}. However, because the experiments 
we are examining contain a regime in which thermal effects play a 
significant role \cite{strick}we must consider them. 

Marko and Siggia \cite{siggia_science,siggia}have derived the free 
energy of a plectoneme in their examination of fluctuations about 
helical structures. To within order unity constants 
\begin{eqnarray}
\label{plect_free_eqn}
E_{pl}&=&\frac{A}{2}\left(\frac{R}{\left(R^2+P^2\right)}\right)^{2}+
\frac{C}{2}\left(\frac{2\pi\sigma_{l}}{h}\right)^2+ 
\\&& \left(\left(R/r_{0}\right)^{-12}
+\left(\pi P/r_{0}\right)^{-12}\right)/r_{0} + \nonumber
\\ &&A^{-1/3} \left(R^{-2/3}+\left(\pi P\right)^{-2/3}\right)\nonumber
\end{eqnarray}
The first two terms in (\ref{plect_free_eqn}) are elastic contributions 
from curvature and twist, respectively. The next line is the hard core 
interaction ($r_0 \simeq 1.75$nm \cite{vologod_r_not}). The last term is 
the entropic penalty incurred for winding too tightly \cite{siggia}. (It is interesting to note that although we include 
the last term in our model, its value is {\em always negligible}.)
Setting the plectonemic fraction X we use (\ref{plect_constr_eqn}) 
and (\ref{plect_free_eqn}) to minimize $E_{pl}$ with respect to $R$ and $P$. 

Next let us determine the thermal behavior of the straight-line 
segment. Such behavior for the {\em untwisted} rod has been examined in some detail by Fixman 
and Kovak \cite{fixman}. Eqn. 
 \ref{perturb_eqn} allows us to make use of their results provided 
we replace $F$ with $F^\prime \equiv F-\frac{p^2}{4A}$. Siggia 
provided a valuable summary of their results in an approximate 
interpolation formula \cite{siggia_comment,siggia}. We employ the above 
substitution in Siggia's result to solve for the thermal shortening 
of the straight portion of the molecule. $Z$ is the actual observed 
extension: 
\be
\frac{\left(F-\frac{p^2}{4A}\right)A}{k_{b} T}= 
\left(\left(1-Z/X\right)^{-2}-1\right)+Z/X
\ee

In the final analysis we compute the optimum plectonemic fraction $X$ 
and the extension $Z$ for a given ${\cal L}k$ and $F$. The results are 
plotted in fig. \ref{trans_fig} side by side with experimental 
results \cite{strick}. Because our model is a very 
simple one, and we have made no attemts to compute exact parameters 
(i.e. 'critical winding', etc.) we cannot claim perfect agreement. 
Nonetheless the resemblance is striking. Our model shows the 
shift from purely thermal behavior for very small $\sigma$ to 
a transition completely driven by elastic considerations for 
larger $\sigma$'s and forces.

\section*{acknowledgements}

The authors would like to acknowledge helpful discussions with Professor J.
White, and with Z. Nussinov.

\begin{figure}
\begin{center}
\leavevmode
\end{center}
\caption[]{\label{duality_fig}
A mapping of arclength onto time
renders the equations governing symmetric elastic rods and spinning
tops identical. The top left figure represents our elastic model of
DNA (see figure \ref{model_fig}). The lower right figure is the conventional
way of representing the motion of the top. The locus of the angles of
inclination of the axis is represented as a curve on the unit sphere.}
\end{figure}

\begin{figure}
\begin{center}
\leavevmode
\end{center}
\caption[]{\label{model_fig}The elastic strand is described by a local
coordinate frame, ${\cal L}$. At each point ${\cal L}$ is related to an
unstressed frame, ${\cal L}_0$,
by Euler angles, The molecule is stretched by $F$ and is required to have a
constant ${\cal L}k$.}
\end{figure}

\begin{figure}
\begin{center}
\leavevmode
\end{center}
\caption[]{\label{poly_fig} The behavior of the solutions is determined
by the polynomial $f\left(u\right)=2F/A
\left(u-a\right)\left(u-b\right)\left(u-c\right)$. The solid curve shows
$f\left(u \right)$ for a typical set of parameters. Motion is defined where
$f\left(u \right) \geq 0$ in the range $u_{-} \leq u \leq u_{+} $. The
shaded curve shows a case where $u_{+} > 1$. The dotted curve ($u_{+} = 1$)
shows the set of parameters that minimize the energy given a fixed ${\cal L}k$.
This is the soliton configuration discussed toward the end of the paper.}
\end{figure}

\begin{figure}
\begin{center}
\leavevmode
\end{center}
\caption[]{\label{energy_fig}The writhing family of solutions. The solution
with the lowest energy is the soliton. In the infinite-length limit the
soliton and the twisted line have the same energy.}
\end{figure}

\begin{figure}
\begin{center}
\leavevmode
\end{center}
\caption[]{\label{plect_fig}The extended and plectonemic phase coexist
in the molecule. The plectonemic phase takes up a portion $X$.}
\end{figure}

\begin{figure}[htbp]
\caption[]{\label{trans_fig}A comparison of our predictions and the data
of Strick et al. The families of curves are $\sigma = 0.102,  0.043,  0.031, 
 0.001, \mbox{and}  0$ from top to bottom. The stretching of the untwisted ($\sigma=0$) 
line is purely entropic; $\sigma=0.102$ transition is dominated by elastic 
energy. No attempts have been made to fit the data.}
\end{figure}

\pagebreak

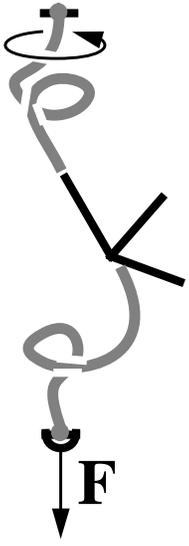
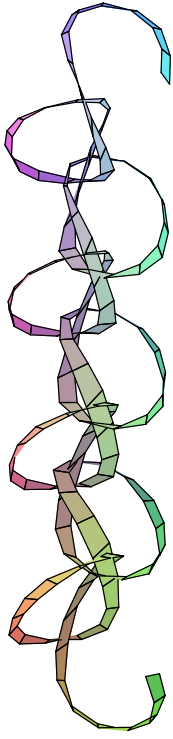
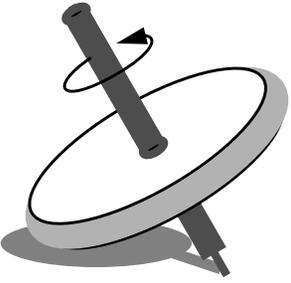
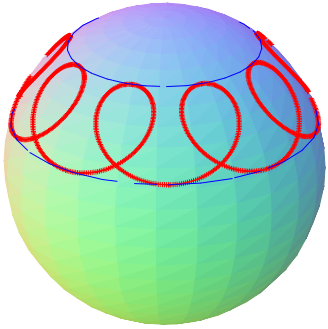
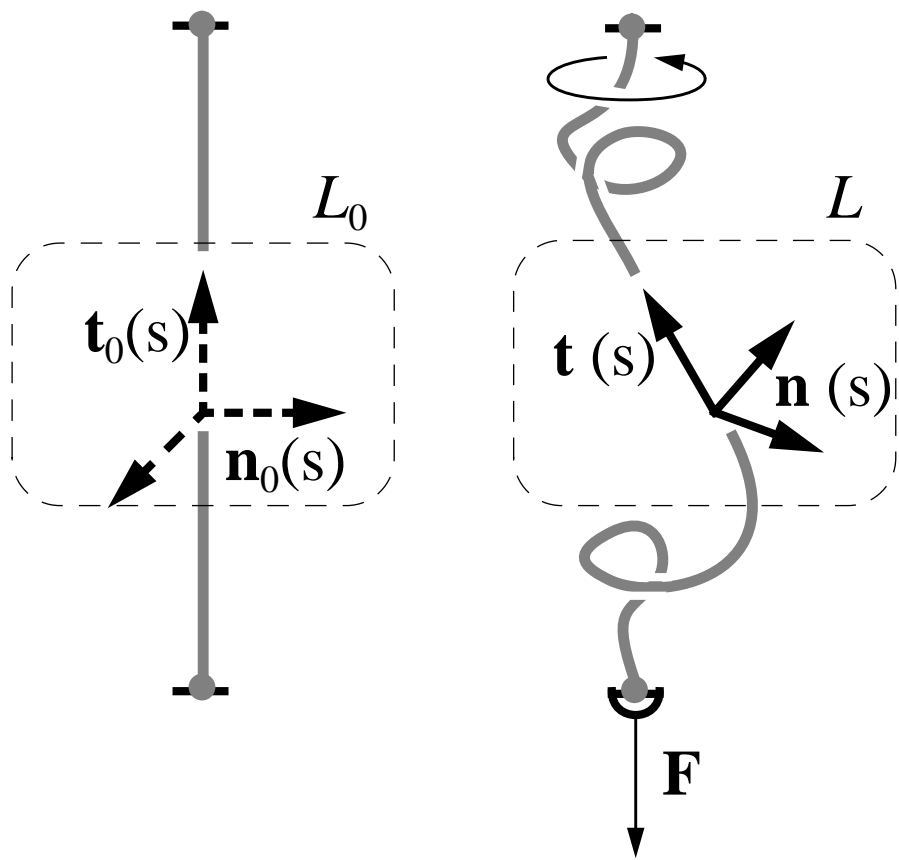
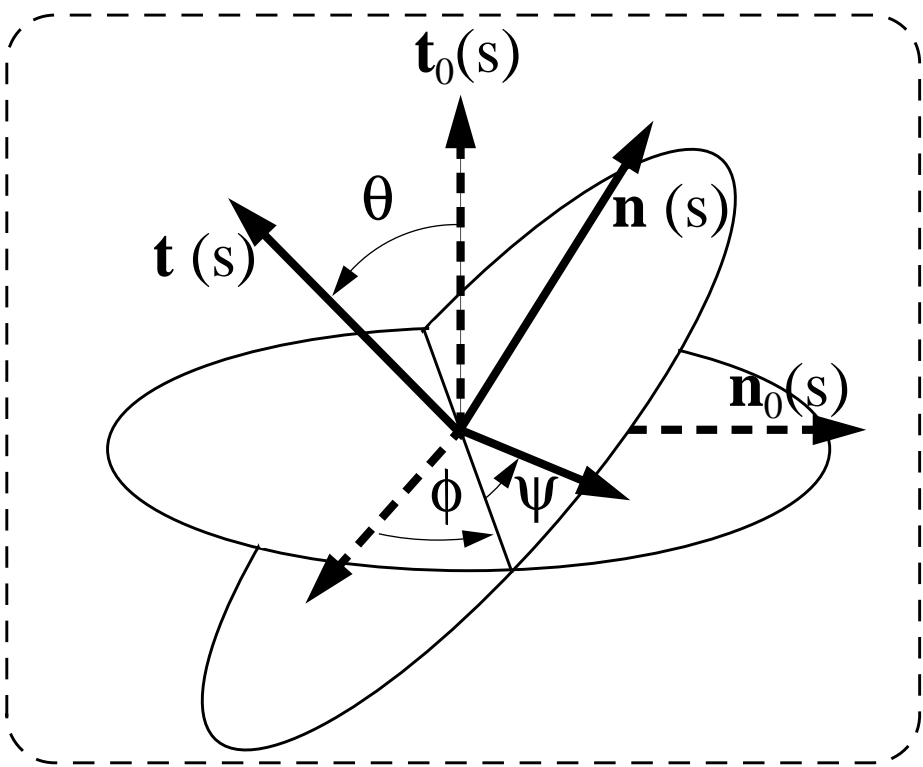
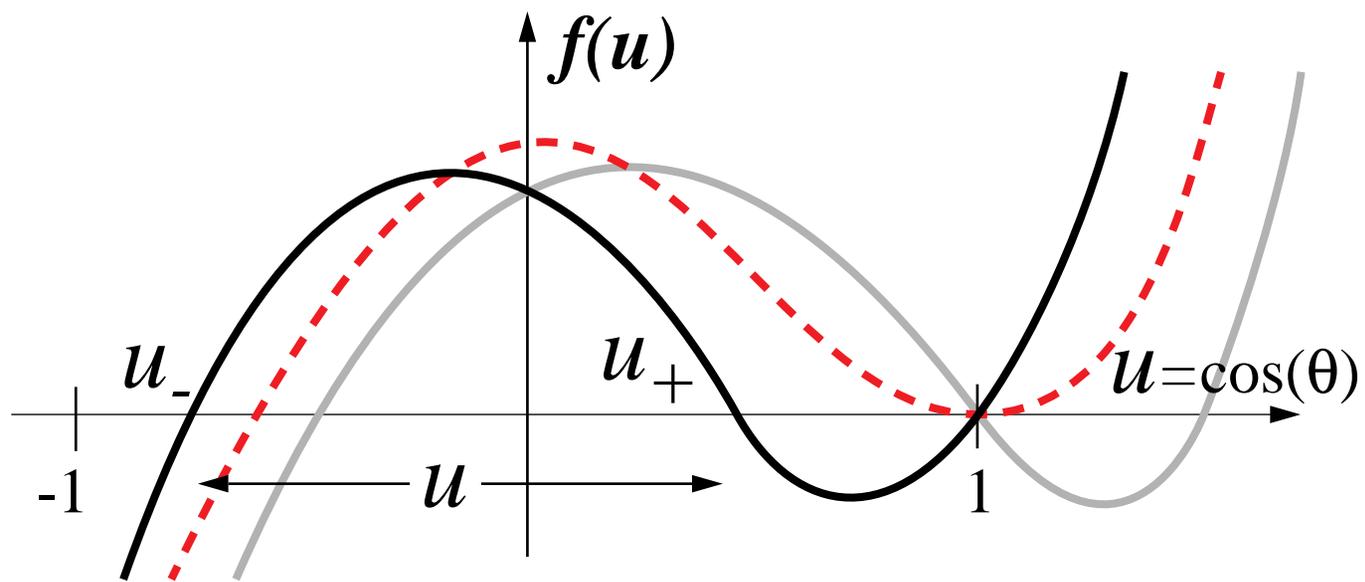
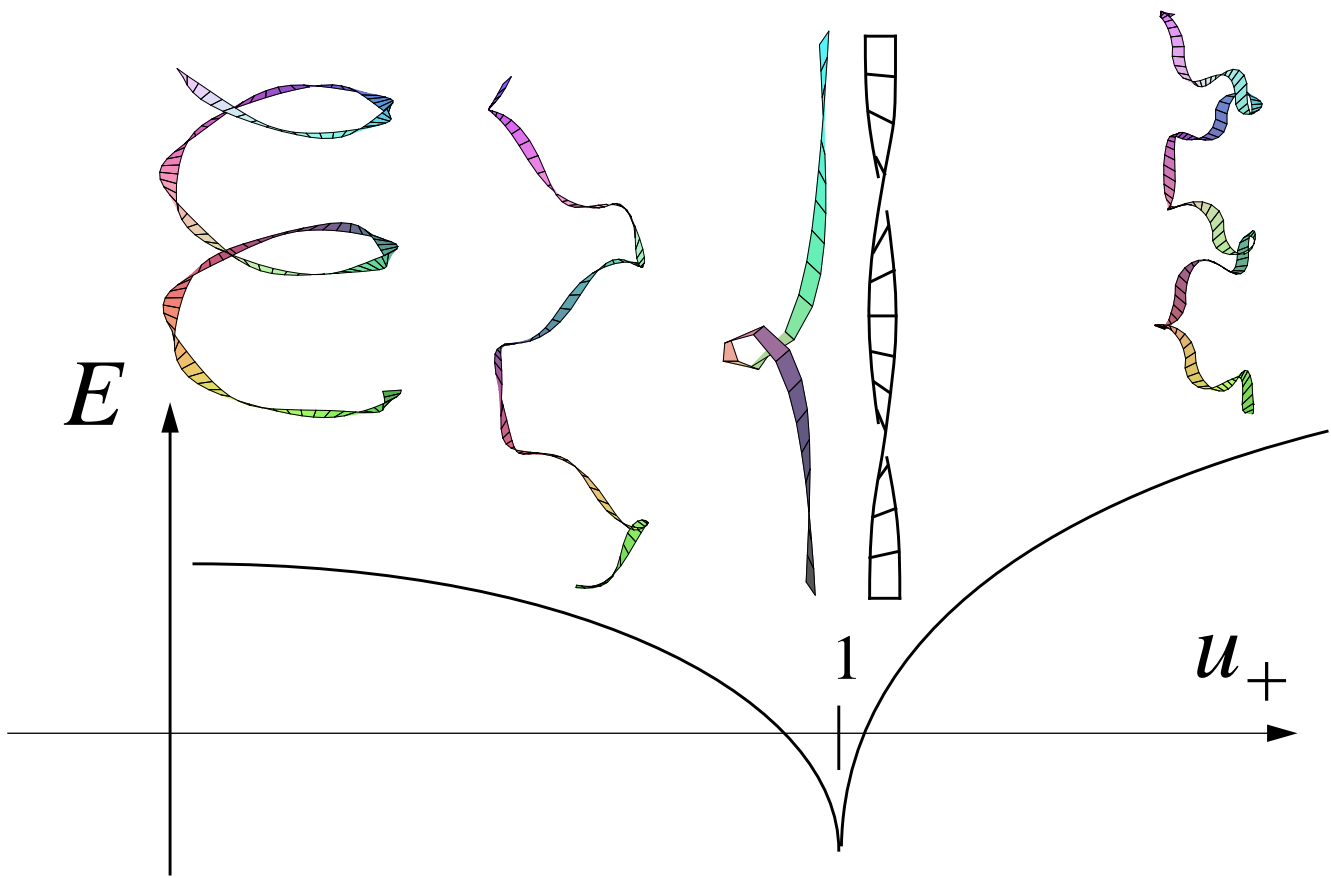
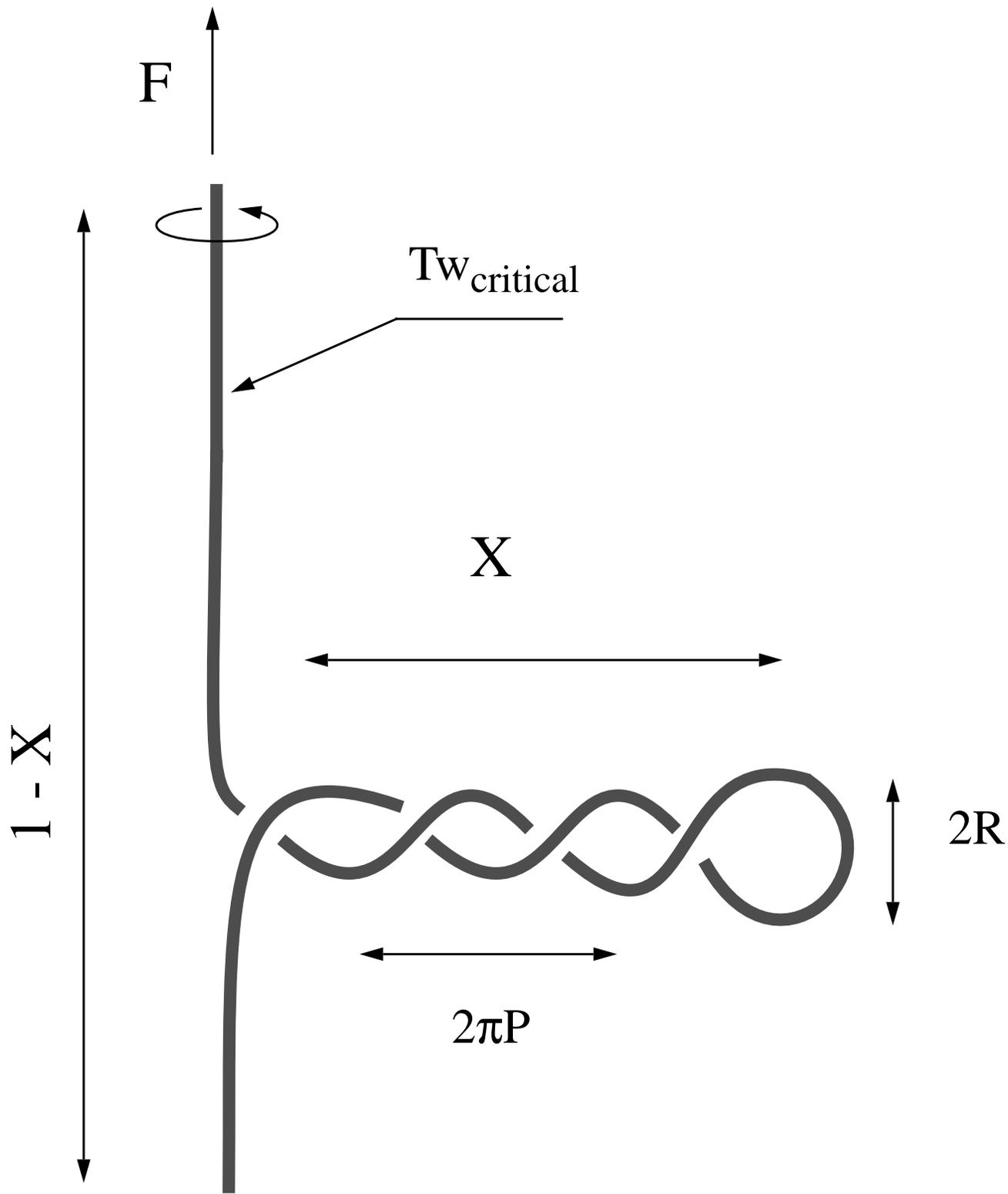
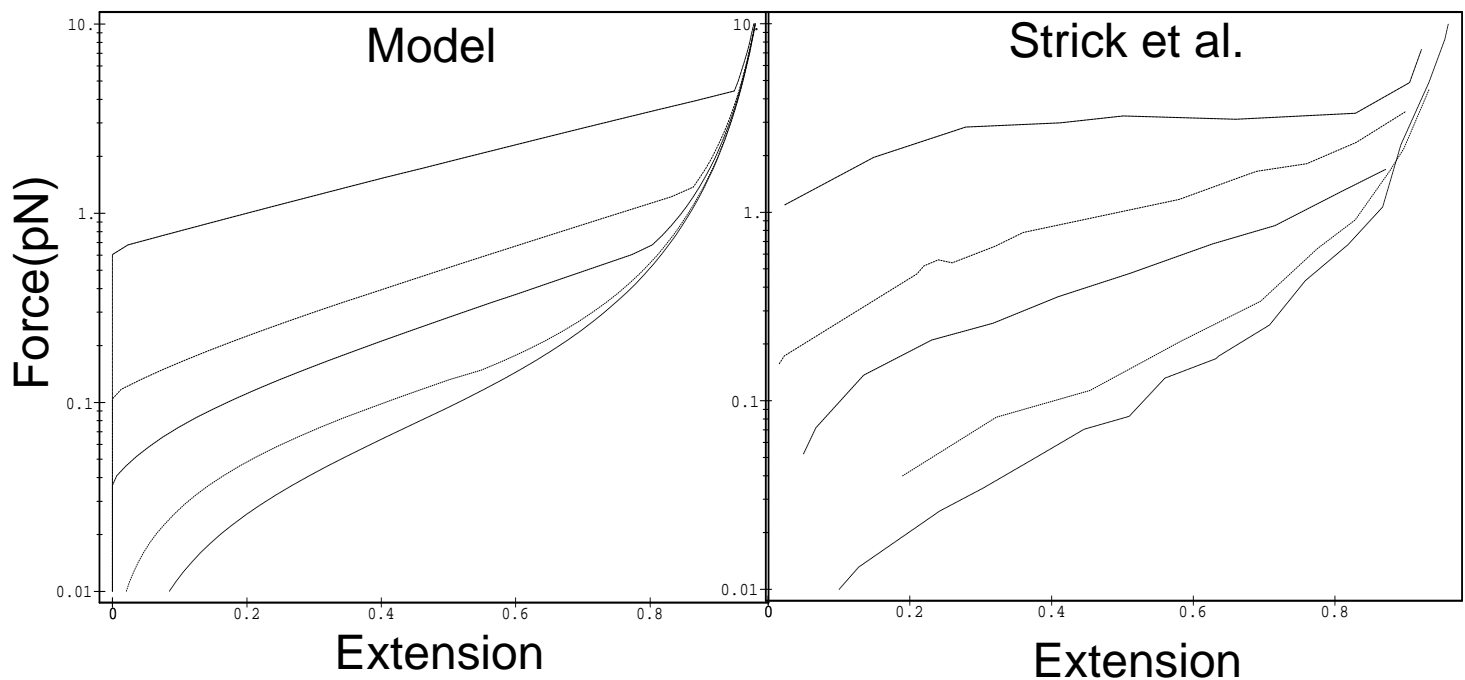
\end{document}